# Data Mining and Analytical Models to Predict and Identify Adverse Drug-drug Interactions

Ricky Wang, Stevenson High School, USA
E-mail: rickywang100@gmail.com

## Abstract

The use of multiple drugs accounts for almost 30% of all hospital admission and is the 5th leading cause of death in America. Since over 30% of all adverse drug events (ADEs) are thought to be caused by drug-drug interactions (DDI), better identification and prediction of administration of known DDIs in primary and secondary care could reduce the number of patients seeking urgent care in hospitals, resulting in substantial savings for health systems worldwide along with better public health. However, current DDI prediction models are prone to confounding biases along with either inaccurate or a lack of access to longitudinal data from Electronic Health Records (EHR) and other drug information such as FDA Adverse Event Reporting System (FAERS) which continue to be the main barriers in measuring the prevalence of DDI and characterizing the phenomenon in medical care. In this review, analytical models including Label Propagation using drug side effect data and Supervised Learning DDI Prediction model using Drug-Gene interactions (DGIs) data are discussed. Improved identification of DDIs in both of these models compared to previous versions are highlighted while limitations that include bias, inaccuracy, and insufficient data are also assessed. A case study of Psoriasis DDI prediction by DGI data using Random Forest Classifier was studied. Transfer Matrix Recurrent Neural Networks (TM-RNN) that address the above limitations are discussed in future works.

## Keywords

# Introduction

Each passing year, more and more adults are being prescribed multiple drugs leading to more and more ADEs, an injury related to a medical intervention. According to the CDC's "HUS 2018 Trend Tables" [1], from 1994 to 2016, the number of people taking three or more prescription drugs has nearly doubled from 11.8% to 21.8% of the population. The increased amount of multiple drug use has introduced a paramount DDI issue. DDI is a phenomenon where one drug increases or decreases the effect of another drug entity. These interactions amongst these drugs may be harmful to the human body. Due to harmful DDI's, the number of emergency hospital visits has dramatically increased [2] causing deaths while ballooning hospital costs. Recently, a new field known as data mining has emerged in order to combat the issues of DDI's. Data mining is about making the tools of data analysis or hypothesis generation catch up with our ability to make those hypotheses. It decodes data from numerous databases and forms hypotheses based on how you train the hypothesis generating model. In the case of DDI's, systems are being built to predict DDIs based on FAERS, EHRs, and other biomedical literatures. Thus, it is essential that not only we build a reliable DDI prediction model, but also a reliable database that can generate accurate and precise data to be extracted from in order to avoid more medical drug accidents.

Unfortunately, previous DDI databases have proven to be very unreliable. While currently 79.7% of hospitals use certified EHRs (Electronic Health Records), and we are hopeful to continue this trend, data before this upsurge has been lost without the consistent use of EHRs in hospitals in the past. Additionally, major confounding biases have been detected throughout these databases making this data even more raw. Due to bias and insufficient and/or a lack of drug information from EHRs and other databases such as the FAERS, even modern examples of DDI prediction models are extremely limited.

There are more advanced supervised machine learning models that have made noticeable improvements in identifying DDIs compared to previous models, yet are consistently limited by bias and lack of adequate data. The first of these models is a label propagation model that performs better than nearest neighbor models used in the past. It also utilizes drug side effects showing significant improvement over previous uses of chemical structures. The second of these models utilizes DGI's (Drug-Gene Interactions) to predict DDI's with machine learning workflow that extracts DDI information from biomedical literatures. Although performance of DDI classifiers were better in both studies, it's hard to say how accurate the prediction models really are if the drug data they are extracting is filled with confounding bias and insufficient data (false positives, lack of data, etc).

While these limitations cannot be totally prevented, we can try to maximize the effectiveness of DDI classifiers using deep learning. Conventional supervised machine learning methods are prone to unattainable automation processes in feature extraction and vocabulary gap, so researchers have proposed a futuristic model using TM-RNN with LSTM (Transfer Matrix Recurrent Network with Long-Short Term Memory) and a memory network. This type of model uses word embeddings based on DDI Extraction 2013 which extracts sentences related to DDIs from biomedical literature effectively eliminating inaccurate data that databases such as FAERS may contain. Additional benefits include: Insensitivity to noise, avoidance of feature engineering, avoidance of overfitting.

# New Advanced Supervised Learning Method One: Label Propagation Based on Clinical Side Effects-2015

Label Propagation is a similarity based approach supervised machine learning method that uses DDI and side effect information from FAERS[3], TWOSIDES[4], SIDER[5], and OFFSIDES[6] to form a network of drugs where certain nodes are confirmed positive DDI. The confirmed DDI's can be used to estimate unknown DDI's within the other drug nodes based on similarity measures such as chemical structures and side effects.

Figure 1. illustrates the difference between k-nearest neighbor method and label propagation. The nearest neighbor method only highlights similar drugs to the training data. Label propagation expands this even further, identifying similar drugs across the entire drug network based on the similar drugs of the training data, the similar drugs of the similar drugs of the training data and so on.

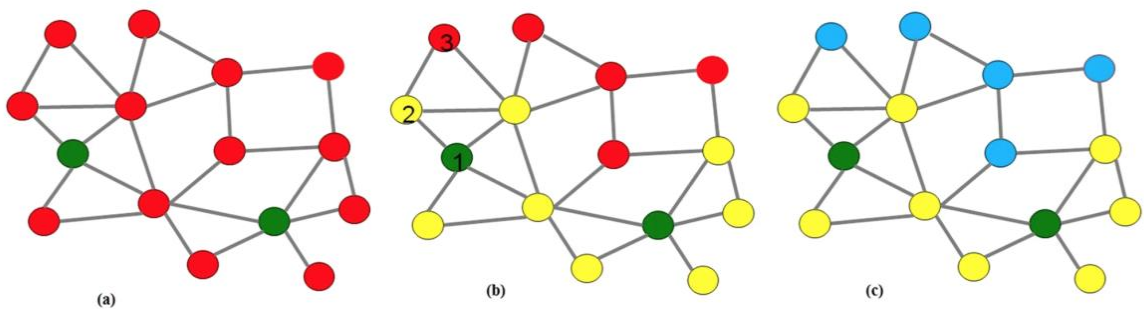

**Figure 1.** Nearest Neighbor Model vs Label Propagation model **[7]** Nearest neighbor model is illustrated in (a) (b) where the model only searches for yellow nodes which are similar to the green ones. Label propagation is illustrated in (c) where the model searches the entire network (yellow + blue nodes)

Researchers used cross validation to build these drug networks dividing the DDI dataset into training and testing subsets. The testing set was chosen with random percentages in the dataset of 15%, 25%, 50%, 75%, 85% (Table 1 and Table 2) and the rest of the data was used for the training set. Before testing, model parameters for the fit curve were tuned using 5-fold cross validation of the training set. ROC and PR curves were created using these different data percentages.

Researchers compared the following Supervised Machine Learning Methods

1. Nearest Neighbor with Chemical Similarity
2. Nearest Neighbor with Label Side Effect Similarity
3. Nearest Neighbor with Off-label Side Effect Similarity
4. Label Propagation with Chemical Similarity
5. Label Propagation with Label Side effect Similarity
6. Label Propagation with Off-label Side Effect Similarity
7. Label Propagation by Integrating All Similarities

Results from Zhang and Wang show that label propagation is far more effective based on the ROC and AUPR scores illustrated in Table 1 and Table 2.

| Methods | 15% | 25% | 50% | 75% | 85% |
| --- | --- | --- | --- | --- | --- |
| NN-Chemical | 0.6951+/−0.0031 | 0.6949+/−0.0021 | 0.6910+/−0.0016 | 0.6838+/−0.0017 | 0.6805+/−0.0017 |
| NN-LabelSE | 0.7359+/−0.0018 | 0.7348+/−0.0019 | 0.7304+/−0.0016 | 0.7288+/−0.0016 | 0.7241+/−0.0020 |
| NN-OffLabelSE | 0.7557+/−0.0017 | 0.7500+/−0.0023 | 0.7436+/−0.0015 | 0.7397+/−0.0015 | 0.7390+/−0.0019 |
| LP-Chemical | 0.8676+/−0.0015 | 0.8654+/−0.0012 | 0.8603+/−0.0007 | 0.8515+/−0.0009 | 0.8415+/−0.0015 |
| LP-LabelSE | 0.8907+/−0.0014 | 0.8880+/−0.0010 | 0.8816+/−0.0007 | 0.8713+/−0.0007 | 0.8642+/−0.0014 |
| LP-OffLabelSE | 0.9219+/−0.0012 | 0.9194+/−0.0010 | 0.9115+/−0.0009 | 0.8994+/−0.0009 | 0.8888+/−0.0012 |
| LP-AllSim | 0.9258+/−0.0011 | 0.9233+/−0.0009 | 0.9156+/−0.0008 | 0.9033+/−0.0008 | 0.8921+/−0.0012 |

**Table 1**. AUROC scores of different methods with different sample sizes show that the LP method has a higher ROC score than NN method regardless of testing sample size [7].

| Methods | 15% | 25% | 50% | 75% | 85% |
|---|---|---|---|---|---|
| NN-Chemical | 0.5182+/−0.0018 | 0.4932+/−0.0020 | 0.4067+/−0.0015 | 0.3810+/−0.0013 | 0.3667+/−0.0014 |
| NN-LabelSE | 0.5537+/−0.0024 | 0.5137+/−0.0019 | 0.4549+/−0.0016 | 0.4038+/−0.0016 | 0.3791+/−0.0015 |
| NN-OffLabelSE | 0.5780+/−0.0022 | 0.5412+/−0.0026 | 0.4710+/−0.0015 | 0.4331+/−0.0016 | 0.3924+/−0.0014 |
| LP-Chemical | 0.6128+/−0.0040 | 0.6026+/−0.0026 | 0.5876+/−0.0016 | 0.5545+/−0.0022 | 0.4726+/−0.0025 |
| LP-LabelSE | 0.6567+/−0.0035 | 0.6481+/−0.0021 | 0.6066+/−0.0019 | 0.5826+/−0.0023 | 0.5323+/−0.0028 |
| LP-OffLabelSE | 0.7195+/−0.0031 | 0.7189+/−0.0019 | 0.6827+/−0.0022 | 0.6477+/−0.0026 | 0.6349+/−0.0031 |
| LP-AllSim | 0.7292+/−0.0032 | 0.7282+/−0.0021 | 0.7052+/−0.0022 | 0.6736+/−0.0025 | 0.6501+/−0.0034 |

**Table 2**. AUPR scores of different methods with different sample sizes show that as testing percent increases, PR scores fall but in general, LP method still has higher AUPR scores than NN methods [7].

ROC curves measure the true positive rate (y axis) against the false positive rate (x axis), thus, the more area under the curve the better the model. As shown in Table 1 and Table 2, while ROC scores are similar throughout each of the testing experiments, the AUPR scores fall off dramatically for each different model as the testing percent (sample size) increases. AUPR provides a quantitative assessment of how well, on average, predicted scores of true interactions are separated from predicted scores of true non-interactions. Thus it holds biological significance in practice: among the best ranked predictions that could potentially be experimentally tested, what proportion of true positives is present. Of course, the larger the sample size, the more false positives there will be explaining the slight drop in AUPR scores for each model.

Additionally, researchers focused on similarity measures based on drug side effect profiles instead of the usual chemical structures. Using side effects has the potential to serve as biomarkers for both therapeutic effects and toxic effects including DDIs since they can be directly observed from humans instead of being simulated in computer models. For example DDI predictions were analyzed between Angiotensin-Converting Enzyme (ACE) Inhibitors (e.g., benazepril, lisinopril and ramipril)/Angiotensin II Receptor Blockers (ARBs)) and nonsteroidal anti-inflammatory drugs (NSAIDs) in which it was known that NSAIDs inhibit prostaglandin-mediated vasodilation and promote salt and water retention. Both of these mechanisms contributing to NSAIDs partially reverse the effects of ACE and ARBs, whose mechanism depends on controlling prostaglandins, renin, or sodium and water balance. The side effects of these drugs can be clearly observed either directly from human reactions or located in the body.

Using the Label Propagation by Integrating All Similarities, results of Zhang and Wang show that off label side effects are the best at predicting DDI's followed by label side effects. Chemical structures are the worst at predicting DDI's as shown in Table 3.

| Sources | 15% | 25% | 50% | 75% | 85% |
|---|---|---|---|---|---|
| chemical structure | 0.2451+/-0.0035 | 0.1417+/-0.0021 | 0.0264+/-0.0012 | 0.0000+/-0.0003 | 0.0000+/-0.0000 |
| label side effect | 0.3412+/-0.0047 | 0.3464+/-0.0018 | 0.3609+/-0.0012 | 0.3505+/-0.0009 | 0.2947+/-0.0008 |
| off-label side effect | 0.4137+/-0.0041 | 0.5119+/-0.0019 | 0.6127+/-0.0016 | 0.6495+/-0.0006 | 0.7055+/-0.0010 |

**Table 3**. Weight vector of features used by the LP method shows that the higher the weight percentage, the more the DDI prediction model depends on that feature. For example, with 85% of the data, off label side effects have a weight of 70.55% while chemical structures have no weight at all demonstrating the higher dependence on off label side effects than any other feature [7]

Still, the model has many limitations: Side effects are often unavailable for clinical candidates, the data from TWOSIDES is derived from FAERS which contains many false positives, and the model predicts "interactions" only and doesn't provide reasoning for why they interact.

# New Advanced Supervised Learning Method Two: Supervised Learning DDI Prediction using Drug Gene Interactions (DGIs)-2017

Since DDIs may occur when two drugs interact with the same gene or when one drug inhibits or induces the metabolic pathway of the other drug, it has also been suggested that incorporating drug-gene interactions (DGIs) can enhance the prediction of DDIs

Researchers wanted to prove that using DGI data along with DDI data will improve the DDI prediction models.

DGI data was extracted from the Comparative Toxicogenomics Database (CTD)[8]. For each drug pair, researchers searched for gene(s) from the CTD database that interacts with both the drugs and retrieved the DGI associations. The process obtained 193,294 DGIs for 5,773 drug pairs (out of 31,268 annotated drug pairs) in DDI Corpus [9], and 49,188 DGIs for 935 drug pairs in MedLine [10] sentences.
The DDI corpus was extracted from MedLine [10] abstracts and DrugBank [12]

UMLS Metathesaurus [11], DrugBank [12] and PharmGKB [13], all biomedical literature databases, were used to compile chemical and drug lexicon.

Researchers Raja and Patrick and Elder used a literature based mining framework to identify chemicals/drugs from a drug/chemical lexicon. Biomedical literature based mining frameworks for DDIs have been gaining popularity recently because biomedical literatures are thoroughly researched by translational scientists making their data more accurate and sufficient for predicting DDIs.

| Dataset | | True Positive | False Positive | False Negative | FP1 | Precision | Recall | F-score | P1 | F1 |
|---|---|---|---|---|---|---|---|---|---|---|
| Training (Cross Validation) | Drug Bank | 11,051 | 2,060 | 932 | 373 | 0.84 | 0.92 | 0.88 | 0.97 | 0.94 |
| | Medline | 1,372 | 484 | 335 | 6 | 0.74 | 0.8 | 0.77 | 1.00 | 0.89 |
| | Overall | 12,423 | 2,544 | 1,267 | 379 | 0.83 | 0.91 | 0.87 | 0.97 | 0.94 |
| Test | Drug Bank | 279 | 61 | 17 | 46 | 0.82 | 0.94 | 0.88 | 0.86 | 0.90 |
| | Medline | 288 | 191 | 58 | 34 | 0.60 | 0.83 | 0.7 | 0.89 | 0.86 |
| | Overall | 567 | 252 | 75 | 80 | 0.69 | 0.88 | 0.78 | 0.88 | 0.88 |

**Table 4**. Accuracy of the lexicons for predicting DDIs using different data sources [14]

To measure the accuracy of the lexicons, F1 score, a score of overall precision (positive predictive value) and recall (model sensitivity), was measured. The overall F1 score for training is 0.87 and 0.78 for testing data shown in Table 4. The accuracy isn't bad but definitely has room for improvement.

In this study, researchers tested many different supervised machine learning classifiers: Bayesian network, logistic regression, support vector machines, decision tree, random tree, random forest, k-nearest neighbors and multilayer perceptron using 10-fold cross validation.

Researchers classified the sentences with potential DDI information, then they classified DDIs to four ADR types:
(*i*) adverse effect
(*ii*) effect at molecular level
(*iii*) effect related to pharmacokinetics
(*iv*) drug interactions without known ADR

According to table 5, compared to DDI features themselves, the addition of DGI features slightly improves each of these classifiers. It should be noted however, that testing DGI features individually dramatically reduces the quality of the DDI prediction model so we should not solely rely on DGI information.

| Classifier | DDI Features | | | DDI and DGI Features | | | DGI Features | | |
|---|---|---|---|---|---|---|---|---|---|
| | Precision | Recall | F-score | Precision | Recall | F-score | Precision | Recall | F-score |
| Bayesian network | 0.93 | 0.69 | 0.79 | 0.93 | 0.69 | 0.79 | 0.54 | 1.00 | 0.71 |
| Decision tree | 0.98 | 0.63 | 0.76 | 0.83 | 0.72 | 0.77 | 0.62 | 0.61 | 0.62 |
| Random tree | 0.76 | 0.77 | 0.76 | 0.79 | 0.77 | 0.78 | 0.69 | 0.71 | 0.70 |

| | | | | | | | | |
|---|---|---|---|---|---|---|---|---|
| Random forest | 0.82 | 0.78 | 0.80 | 0.84 | 0.78 | 0.81 | 0.70 | 0.71 | 0.70 |
| K-nearest neighbors | 0.76 | 0.73 | 0.74 | 0.76 | 0.77 | 0.76 | 0.69 | 0.73 | 0.71 |

Table 5. Comparison of DDI, DDI/DGI and DGI with different supervised machine learning methods [14]

# Psoriasis Case Study

Psoriasis is among the fifteen diseases identified to pose significant socioeconomic and public health burden. In the United States, the cost spent on psoriasis is estimated to be approximately $112 billion dollars annually.

Three drugs methotrexate, cholecalciferol and mycophenolic acid are suggested for psoriasis and were identified to undergo DDI's based on DGI information. By incorporating the DGI information, the researchers' approach can predict DDIs and ADRs if the drug pairs are not present in the same sentence from the literature.

From Figure 2 of Raja, Patrick and Elder, 31 DDI's (11 for adverse effect; 10 for effect at molecular level; and 10 for effect related to pharmacokinetics) are associated with methotrexate, cholecalciferol and mycophenolic acid that were not identified in the same sentence from the literature

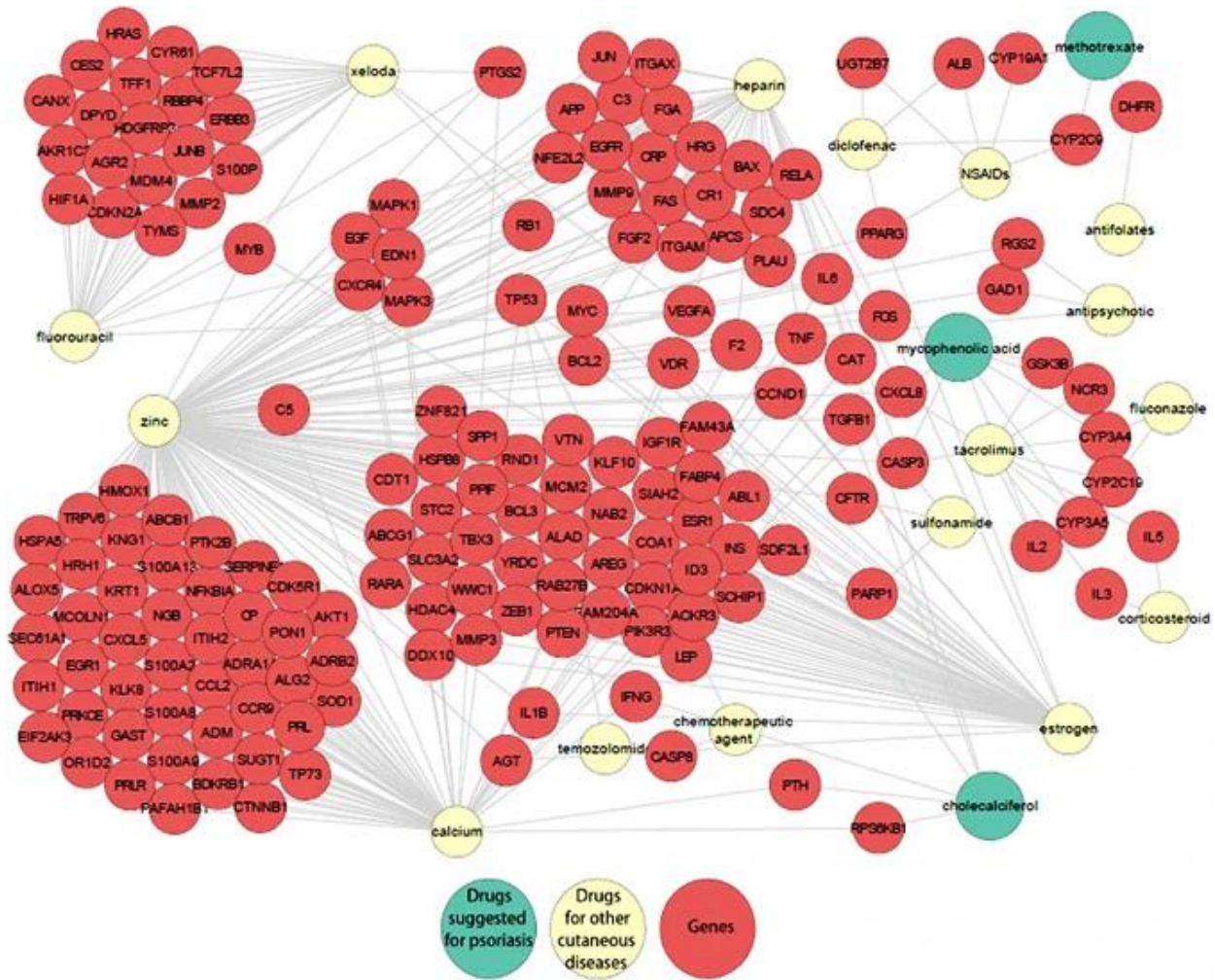

**Figure 2.** Random forest classifier based on DGI information for Psoriasis DDI prediction study [14].

As illustrated in Figure 2, drugs suggested for psoriasis were implemented into a random forest classifier with other drugs known to treat cutaneous diseases. This drug-gene network shows the interactions of drugs suggested for psoriasis and drugs for other cutaneous diseases with similar genes suggesting that there could be ADRs associated between the two types of drugs which were confirmed after random samples of the predictions were traced back to PubMed databases.

Limitations: CTD has not been used to predict DGI's in common practice. The CTD may also have false reports as well and unknown DGI's are obviously not included. Among 31,268 annotated drug pairs from 3,788 sentences in the DDI corpus, only 5,773 drug pairs from 1,441 sentences contain DGI information.

Like any database, there are always false positives clouding the accuracy of the prediction model. Removing these false positive drugs increase the F1-score of the lexicons to 0.94 for training and 0.88 for testing compared to 0.87 and 0.78 respectively without their removal. Chemical and drug lexicon of course contain false negatives as well (i.e. Brand names, partial identification of drug names, drug classes that aren't chemicals). Unified Medical Language System National Drug File – Reference Terminology [11] was added to the lexicon in order to confirm that drugs in the lexicon are actually chemicals.

## Overarching Limitations

Although the Label Propagation model using drug side effects and Supervised Learning DDI Prediction model using Drug-Gene interactions(DGIs) has better DDI prediction accuracy than earlier models, neither of them are perfect and both share limitations, particularly confounding bias. Confounding bias is often a problem for novel DDI detection algorithms. For example, DDI predictions only take into account the drugs themselves and do not consider increased/decreased risk within certain genders, ages, ethnicities, cities, etc.

A recent study taken in Blumenau, Brazil by Brattig, de Araujo Kohler, Mattos, and Rocha [15] exposes these biases that should inform drug physicians and pharmacologists that these confounding biases shouldn't be considered lightly when managing their carousel of drugs. Those more at risk should be more accounted for and better protected from adverse drug events.

It was reported that in Blumenau, 10,734 females (69.46%) and 4793 males (30.54%) were co-administered drugs. Of course, the greater number of co-administration for females should correspond to a greater number of DDI's for females. This effect was mitigated during the studies, however, where the researchers measured risk of co-administration (RC) compared to risk of getting a DDI (RI). Theoretically if there was not any significant gender bias, the RC and RI would match. In the study though, the RI was around 1.6 compared to only 1 for the RC. This gender bias was further exposed using a drug network measured according to patient volume. Out of the 181 drug networks, 133 pose a greater risk for females. This phenomenon could be explained by more prescriptions for female hormone therapy drugs which would only target females, however, the results do not change much when these drugs are removed as women still account for 116/158 networks as illustrated in Figure 3.

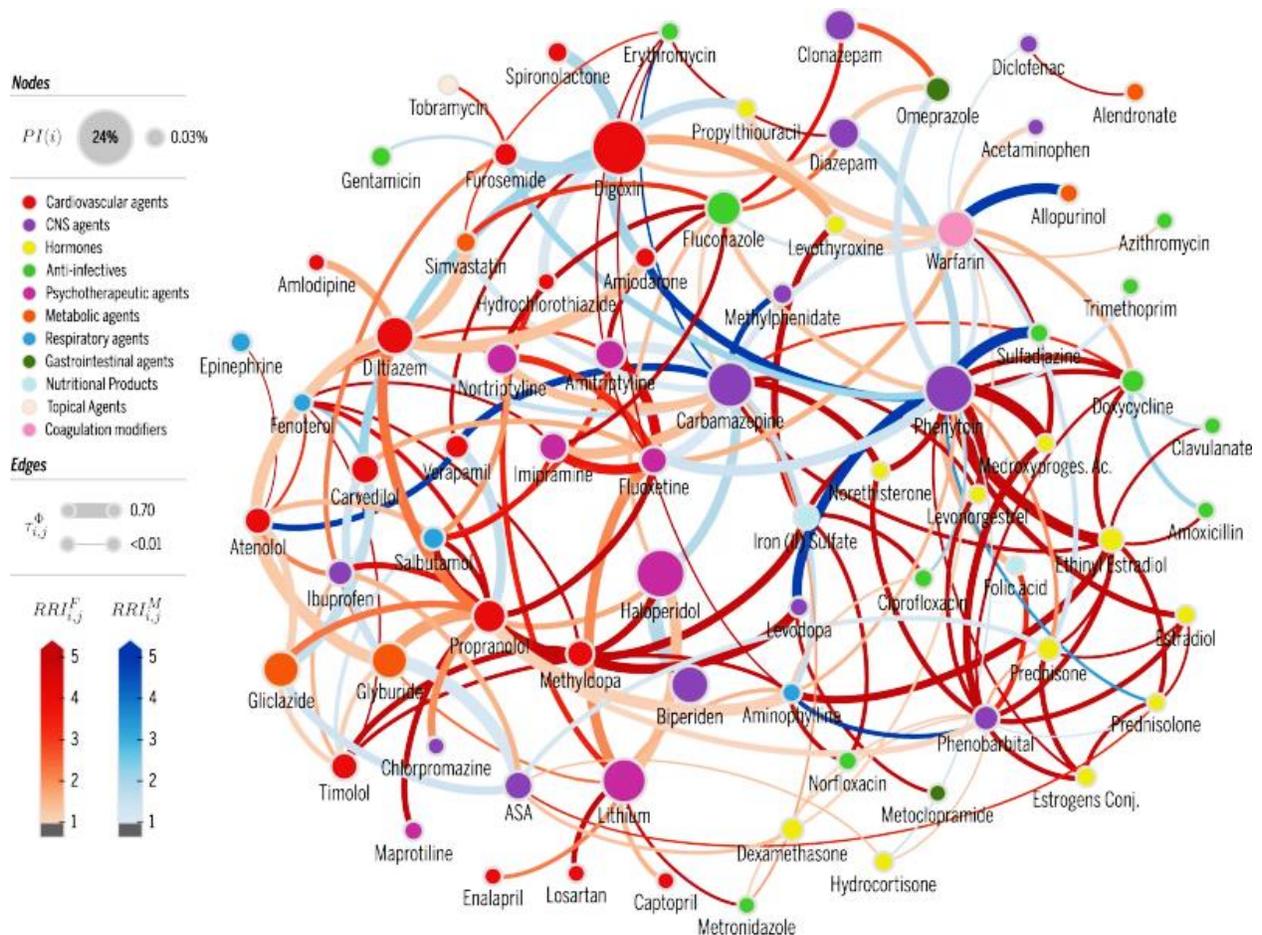

**Figure 3.** Blumenau City-wide electronic health records reveal gender and age biases in administration of known drug–drug interactions (blue= men at risk, red=women at risk) where women hold significantly more risk of DDIs [15].

An increase in age follows a similar pattern in DDI prediction models. RI varies significantly more compared to other age groups than does the variance in RC. To further investigate age bias, an unbiased statistical null model was developed by Brattig, de Araujo Kohler, Mattos, and Rocha in Figure 4 to model RC and RI. The RC curves greatly resembled the null model while the RI curve had a larger RI for older ages than the null model.

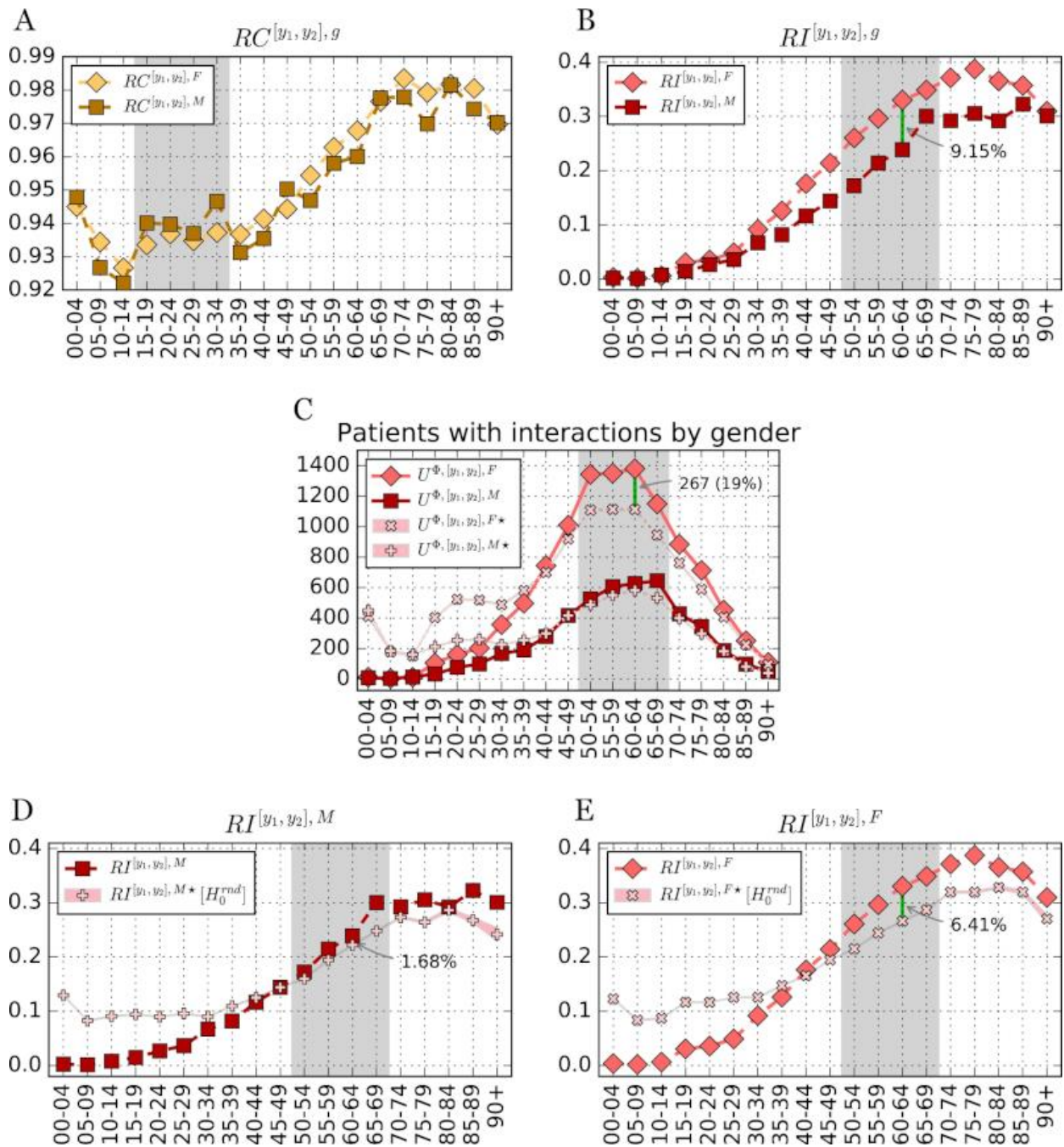

**Figure 4.** (star= null model) Risk of co-administration and interaction per age range and gender. (a) Risk of co-administration per age group and gender. (b) Risk of interaction per age group and gender. (c) Absolute number of patients with at least one known DDI co-administration, per age and gender. (d) (e) Female and male risk of interaction per age group and gender [15].

These biases suggest that DDI data can be flawed in numerous ways and there still needs to be a lot more work done with these DDI models before physicians and pharmacologists can receive reliable DDI info for clinical use.

## Future Works: Deep Learning Neural Networks (TM-RNN)-2020

It's essential that we prepare medical centers with reliable information about DDI's in order to prevent future adverse drug events from occurring in patients. While there are rules based and supervised machine learning models that readily store information about DDI's, these models are susceptible to issues such as a "vocabulary gap" and unattainable automation process in feature extraction.

A new deep learning biomedical literature based model using Long-Short Term Memory (LSTM), a Transfer Weight Matrix, and a Memory Network has been proposed extracting data from DDIExtraction2013 [16], composed of DrugBank [12] and Medline [10], which is a vocabulary index of biomedical literature on drug-drug interactions.

There are many benefits to TM-RNN. First of all, using biomedical literature instead of databases such as FAERS is a more accurate approach to creating the DDI prediction models since DDIExtraction2013 should contain less false positives than FAERS database. Additionally, this study avoids feature engineering by using neural networks. This model uses deep learning to learn a complex hierarchical representation of the data and generates new latent features. Automated feature engineering has the potential to reduce the amount of confounding bias in these models since if trained properly, the neural network can precisely control each confounding factor.

LSTM with transfer matrix: The addition of LSTM helps clean up the word lexicon by capturing word order and long word dependency information. For example, if Drug A is associated with Drug B, the LSTM remembers that information and if Drug A is imputed into the neural network again, the network can spit out a different interaction of Drug A with another drug instead of spitting out the same interaction with Drug B LSTM with TM can also generate a variety of sentences and structure preventing overfitting in the model. Thus, noise from negative samples is generally prevented.
To prove that the transfer matrix enhances the neural networks performance, researchers tested the model without the TM. The transfer matrix model depicts higher precision, recall and F1 scores compared to the baseline clearly making it the superior model as shown in Table 6.

|  | Baseline | | | TM-RNN | | |
| --- | --- | --- | --- | --- | --- | --- |
|  | P | R | F | P | R | F |
| Int | 88.57 | 32.29 | 47.33 | 82.22 | 38.54 | **52.48** |
| Advice | 81.87 | 63.93 | 71.79 | 81.44 | 72.15 | **76.51** |
| Effect | 64.25 | 69.47 | 66.76 | 69.27 | 71.99 | **70.60** |
| Mechanism | 79.78 | 71.48 | 75.40 | 74.13 | 78.86 | **76.42** |
| Overall | 73.57 | 65.15 | 69.11 | 74.11 | 70.82 | **72.43** |

**Table 6**. Three Statistics of Baseline and TM-RNN (Both Models Use Two-Layer Bi-LSTM) [17]

Feature Fusion with Memory Network: A feature that complements the LSTM transfer matrix is the memory network. The LSTM recurrent neural network requires memory space that allows it to recall previous output results to capture word order and long word dependency information. However, crosstalk between this memory space and input information may introduce unnecessary noise into the system. As a result, researchers created a memory network that prevents crosstalk between the memory space and input information. As shown in Figure 5 by Liu and Huang, instead of directly accessing the memory space for classification, an alternative pathway is used where certain attentive weights are required to access the memory space where it can then proceed towards classification within a SoftMax Classifier. The alternative pathway prevents crosstalk of memory space with the input information .

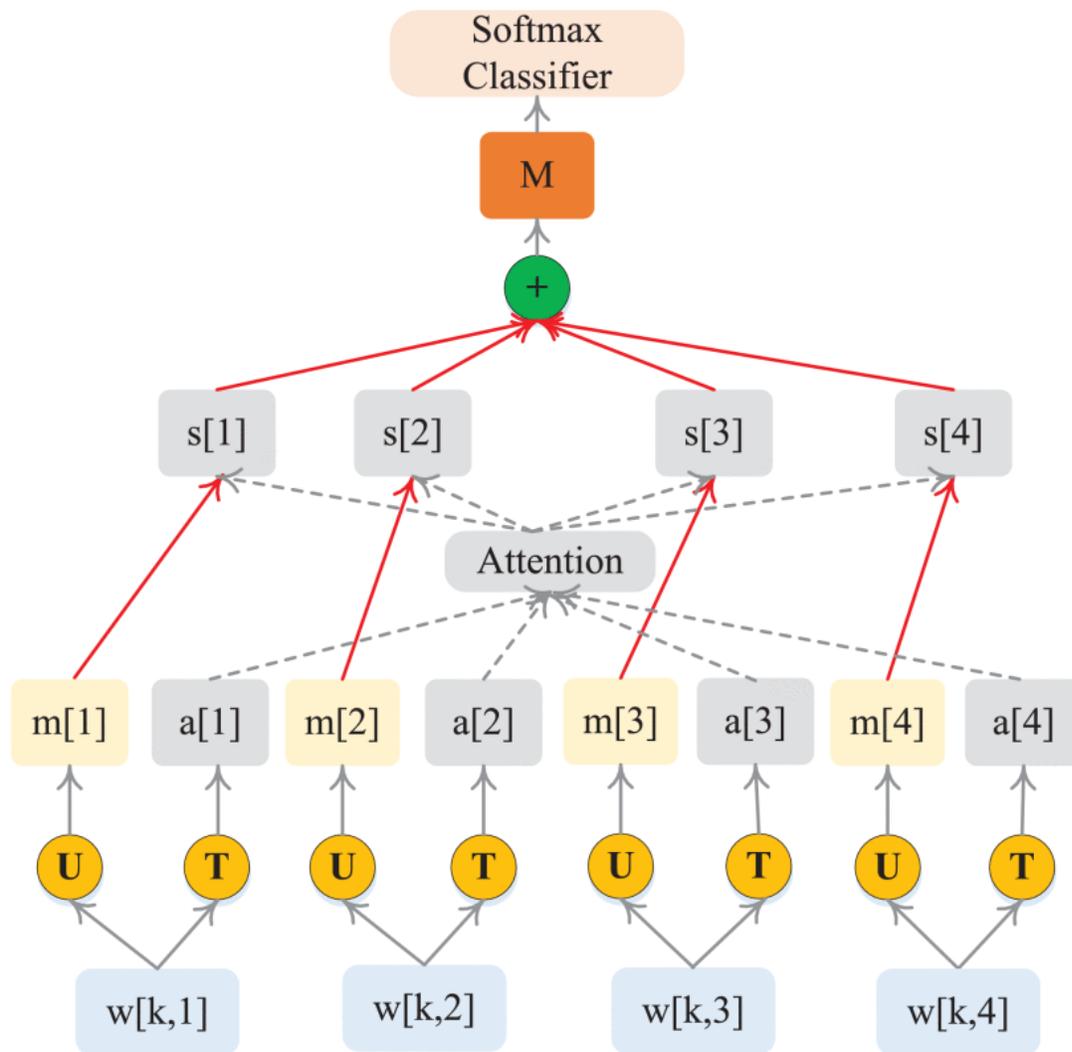

**Figure 5.** Feature fusion with memory network for drug-drug type classification [17]

As expected, this new type of deep learning DDI identification model was deemed superior to other rules based and SML based databases. Deep learning methods completely avoid feature engineering and don't need to go through classification choice problems which results in better precision, recall, and f-score according to Figure 6.

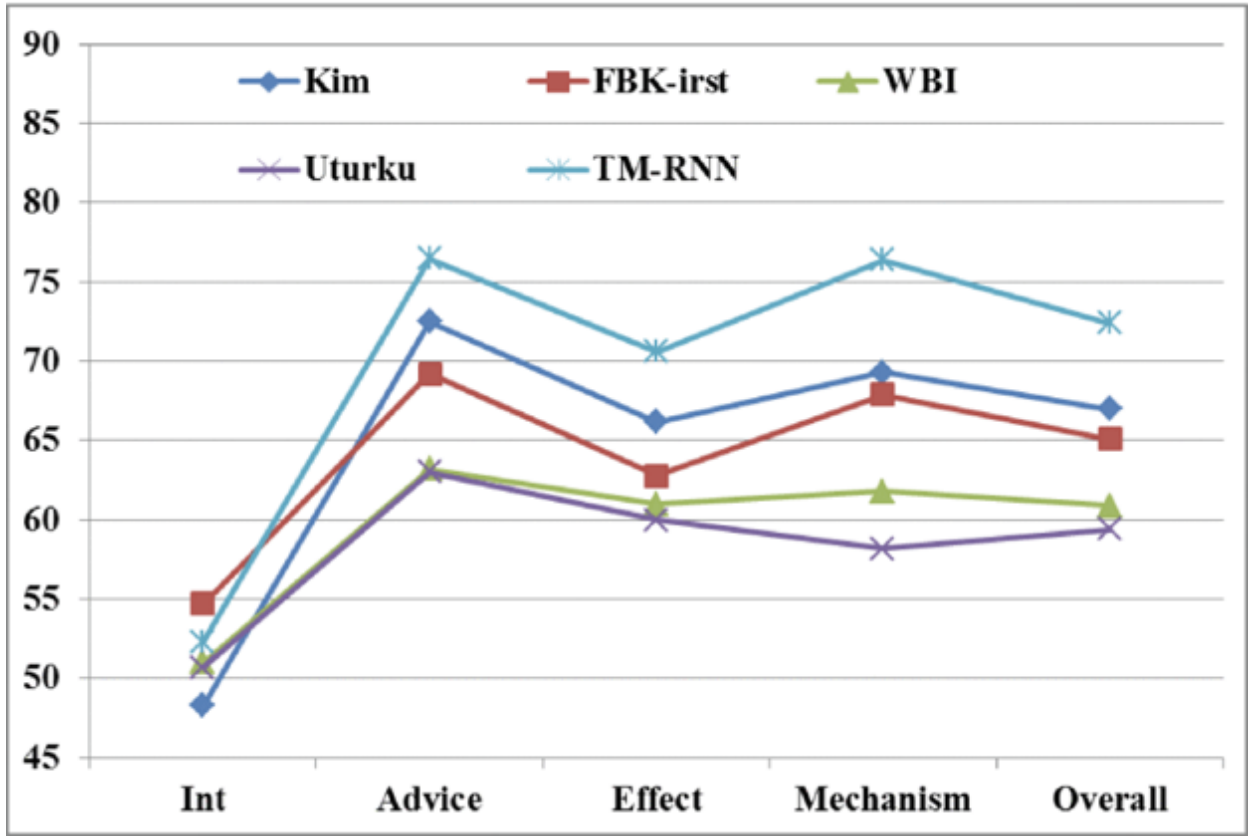

**Figure 6.** Comparison with other conventional SML-based methods. performance of Kim's linear SVM-based method [21], FBK-irst [20], WBI [18] and UTurku [19].

Whereas rules based and SML methods are significantly outperformed by TM-RNN in almost every single category, researchers also effectively compared TM-RNN to previous deep learning models. These deep learning models include convolutional neural networks, and recurrent neural networks without the transfer matrix. According to Figure 7, convolutional and recurrent neural networks achieve better precision compared to TM-RNN. However, TM-RNN achieves much greater recall and an overall higher f-score.

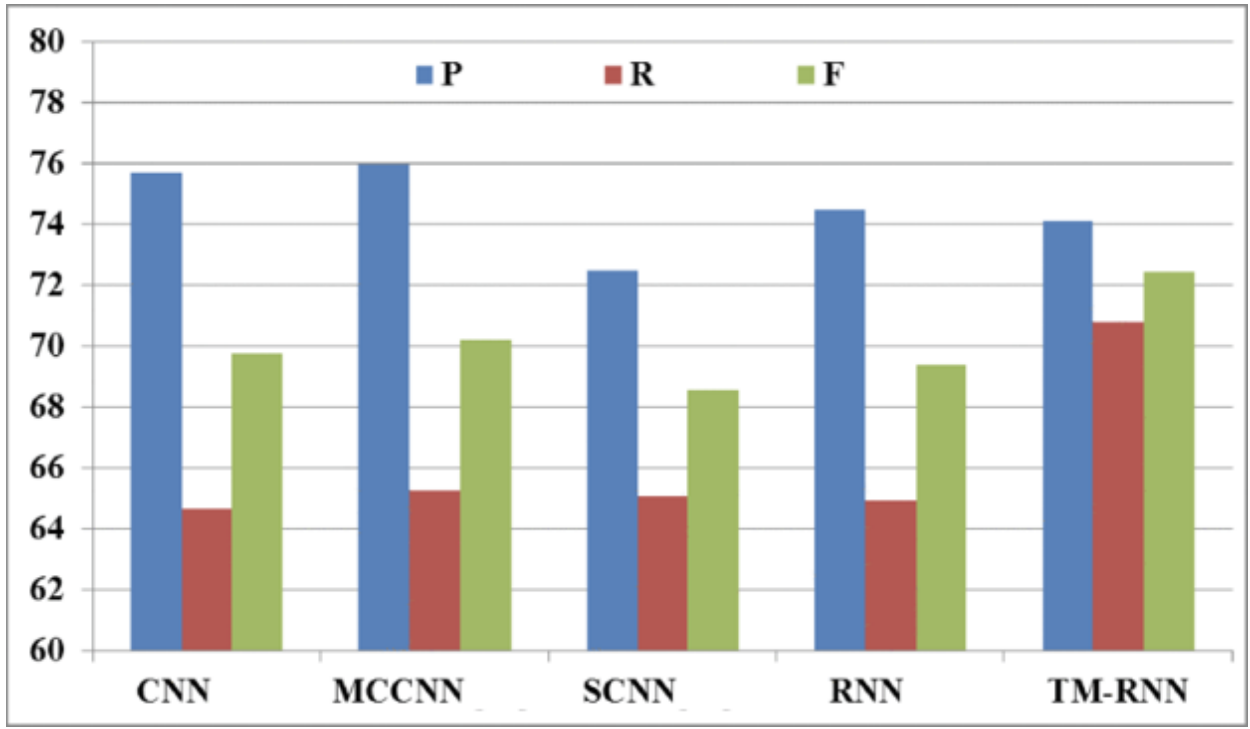

**Figure 7.** Comparisons with other neural networks based methods such as convolutional [22], multichannel convolutional [23], syntax convolutional [24], and recurrent neural networks without transfer matrix [25].

TM-RNN achieves a significantly higher recall score because the LSTM transfer matrix and memory network make it much less sensitive to noise compared to other neural networks. Table 7 illustrates the f-scores of each different neural network before and after preprocessing, a data cleaning method. TM-RNN's scores aren't affected as much by preprocessing (change of 1.61) as the convolutional networks (change of 4.75, 4.1) and the recurrent neural network without TM (change of 2.11). Since preprocessing remains a very tedious step in data mining, it's essential that models are built with the highest initial f-score possible.

| Methods | Before | After | Net Change |
|---|---|---|---|
| CNN | 65.00 | 69.75 | 4.75 |
| MCCNN | 67.80 | 70.21 | 2.41 |
| SCNN | 64.50 | 68.43 | 4.1 |
| RNN | 67.28 | 69.39 | 2.11 |
| TM-RNN | 70.82 | 72.43 | 1.61 |

**Table 7.** Effect of preprocessing on other neural networks based methods such as convolutional [22], multichannel convolutional [23], syntax convolutional [24], and recurrent neural networks without transfer matrix [25].

This method reduces confounding bias, overfitting, and noise from negative samples making it easier for doctors and physicians to prescribe drugs to patients with more knowledge on the possibility of DDIs.

Limitations: DDIExtraction2013 data may not be complete and FAERS data may have more DDI's available. Although the LSTM can encode any sentence length, an extremely long sentence and two target drug entities that are too far from each other will still introduce noise information. A typical error is that the input sample will be misclassified as 'Other', which is equivalent to noise in the datasets.

# Conclusion

Drug-drug interactions are becoming a staple of modern medicine. Accounting for 30% of all ADEs, DDIs have seriously affected patients and ballooned hospital costs. At the same time though, the drastic consequences of DDIs have forced pharmacologists and translational scientists to develop new treatments and vaccines for the public. While development can be facilitated by in vivo experiments, the numerous amounts of DDIs that are possible makes that not plausible. Developing these medicines requires a proper algorithm to simulate DDIs in patients. Over the years, supervised machine learning algorithms have been developed, yet they contain too many limitations for reliable clinical use. As we enter a new decade, there remains a tremendous opportunity to leverage the power of deep learning to break open a new avenue of discovering DDIs effectively saving lives and reducing hospital costs. Unfortunately, deep learning is still a work in progress as we haven't been able to fully articulate the biomedical literature yet. Sentences that are too complex or lengthy could be interpreted incorrectly and introduce unnecessary noise to the system. As scientists continue to improve deep learning methods, they must begin to develop mastery over the semantics of these biomedical literatures. Doing so would result in an extraordinary medical breakthrough moving ever so closer to eliminating the repercussions of DDIs.

# Reference

[1] Rui P, Kang K. *National Hospital Ambulatory Medical Care Survey: 2017 emergency department summary tables*. National Center for Health Statistics, 2017. Available from: https://www.cdc.gov/nchs/data/nhamcs/web_tables/2017_ed_web_tables-508.pdf

[2] NCHS, *National Health and Nutrition Examination Survey.* National Center for Health Statistics, 1988-2016. Available from: https://www.cdc.gov/nchs/data/hus/2018/038.pdf

[3] Sakaeda, Toshiyuki et al. Data mining of the public version of the FDA Adverse Event Reporting System. *International journal of medical sciences.* 2013; vol. 10,7 796-803. Available from: doi:10.7150/ijms.6048

[4] Tatonetti, N. P., Ye, P. P., Daneshjou, R. & Altman, R. B. Data-driven prediction of drug effects and interactions. 2012; *Sci Transl Med 4*, 125ra31. Available from: doi: 10.1126/scitranslmed.3003377-TWOSIDES

[5] Kuhn, Michael et al. The SIDER database of drugs and side effects. *Nucleic acids research.* 2016; vol. 44,D1: D1075-9. Available from: doi:10.1093/nar/gkv1075

[6] Tatonetti, N. P., Ye, P. P., Daneshjou, R. & Altman, R. B. Data-driven prediction of drug effects and interactions. 2012; *Sci Transl Med 4*, 125ra31. Available from: doi: 10.1126/scitranslmed.3003377-OFFSIDES

[7] Zhang, P., Wang, F., Hu, J. *et al.* Label Propagation Prediction of Drug-Drug Interactions Based on Clinical Side Effects. 2015; *Sci Rep* 5, 12339. Available from: https://doi.org/10.1038/srep12339

[8] Davis, A. P. *et al.* The Comparative Toxicogenomics Database's 10th year anniversary: update 2015. *Nucleic acids research* 43, D914–920. Available from: doi:10.1093/nar/gku935

[9] Segura-Bedmar, I., Martinez, P. & Herrero-Zazo, M. Lessons learnt from the DDIExtraction-2013 Shared Task. 2014; *J Biomed Inform* 51, 152–164. Available from: doi:10.1016/j.jbi.2014.05.007- DDI Corpus

[10] Segura-Bedmar, I., Martinez, P. & Herrero-Zazo, M. Lessons learnt from the DDIExtraction-2013 Shared Task. 2014; *J Biomed Inform* 51, 152–164. Available from: doi:10.1016/j.jbi.2014.05.007- MedLine

[11] Bodenreider, Olivier. The Unified Medical Language System (UMLS): integrating biomedical terminology. 2004; *Nucleic acids research* vol. 32, Database issue: D267-70. Available from: doi:10.1093/nar/gkh061

[12] Segura-Bedmar, I., Martinez, P. & Herrero-Zazo, M. Lessons learnt from the DDIExtraction-2013 Shared Task. 2014; *J Biomed Inform* 51, 152–164. Available from: doi:10.1016/j.jbi.2014.05.007- DrugBank

[13] Thorn, Caroline F et al. PharmGKB: the Pharmacogenomics Knowledge Base. 2013; *Methods in molecular biology (Clifton, N.J.)* vol. 1015 : 311-20. Available from: doi:10.1007/978-1-62703-435-7_20

[14] Raja, K., Patrick, M., Elder, J.T. *et al.* Machine learning workflow to enhance predictions of Adverse Drug Reactions (ADRs) through drug-gene interactions: application to drugs for cutaneous diseases. 2017; *Sci Rep* 7, 3690. Available from: https://doi.org/10.1038/s41598-017-03914-3

[15] Brattig Correia, R., de Araújo Kohler, L., Mattos, M.M. *et al*. City-wide electronic health records reveal gender and age biases in administration of known drug–drug interactions. 2019; *npj Digit. Med.* 2, 74. Available from: https://doi.org/10.1038/s41746-019-0141-x

[16] Segura-Bedmar, I., Martinez, P. & Herrero-Zazo, M. Lessons learnt from the DDIExtraction-2013 Shared Task. 2014; *J Biomed Inform* 51, 152–164. Available from: doi:10.1016/j.jbi.2014.05.007- DDI Extraction 2013

[17] J. Liu, Z. Huang, F. Ren and L. Hua, Drug-Drug Interaction Extraction Based on Transfer Weight Matrix and Memory Network. 2019; *in IEEE Access*, vol. 7, pp. 101260-101268. Available from: doi: 10.1109/ACCESS.2019.2930641.

[18] P. Thomas, M. Neves, T. Rocktäschel and U. Leser, WBI-DDI: Drug-drug interaction extraction using majority voting.2013; *Proc. SemEval*, pp. 628-635. Available from: https://www.aclweb.org/anthology/S13-2105.pdf

[19] J. Björne, S. Kaewphan and T. Salakoski, UTurku: Drug named entity recognition and drug-drug interaction extraction using SVM classification and domain knowledge. 2013; *Proc. SemEval*, pp. 651-659. Available from: https://www.aclweb.org/anthology/S13-2108.pdf

[20] M. F. M. Chowdhury and A. Lavelli, FBK-irst: A multi-phase kernel based approach for drug-drug interaction detection and classification that exploits linguistic information. 2013; *Proc. SemEval*, pp. 351-355. Available from: https://www.aclweb.org/anthology/S13-2057.pdf

[21] K. Sun, H. Liu, L. Yeganova and W. J. Wilbur, Extracting drug-drug interactions from literature using a rich feature-based linear kernel approach. 2015; *J. Biomed. Inform.*, vol. 55, pp. 23-30. Available from: https://www.sciencedirect.com/science/article/pii/S1532046415000441

[22] S. Liu, B. Tang, Q. Chen and X. Wang, "Drug-drug interaction extraction via convolutional neural networks". 2016; *Comput. Math. Methods Med.*, vol. 2016, pp. 1-8. Available from: https://www.hindawi.com/journals/cmmm/2016/6918381/

[23] C. Quan, L. Hua, X. Sun and W. Bai, Multichannel convolutional neural network for biological relation extraction. 2016; *BioMed. Res. Int.*, vol. 2016, no. 1, pp. 1-10. Available from: https://www.hindawi.com/journals/bmri/2016/1850404/

[24] Z. Zhao, Z. Yang, L. Luo, H. Lin and J. Wang, Drug drug interaction extraction from biomedical literature using syntax convolutional neural network. 2016; *Bioinformatics*, vol. 32, no. 22, pp. 3444-3453. Available from: https://academic.oup.com/bioinformatics/article/32/22/3444/2525603

[25] S. K. Sahu and A. Anand, Drug-drug interaction extraction from biomedical texts using long short-term memory network. 2018; *J. Biomed. Inform.*, vol. 86, pp. 15-24. Available from: https://www.sciencedirect.com/science/article/pii/S1532046418301606